\documentclass[11pt, a4paper]{article}
\usepackage[T1]{fontenc}
\usepackage{a4wide, url}
\usepackage{graphicx}
\usepackage{natbib}
\usepackage{enumitem}
\usepackage{hyperref}
\usepackage[labelfont=bf]{caption}
\usepackage{setspace}
\usepackage{multirow}
\usepackage{amsmath, amsthm, amssymb, amsfonts}
\usepackage{bm}
\usepackage{multirow}
\usepackage{color}     
\usepackage[ruled, vlined]{algorithm2e}

\allowdisplaybreaks

\DeclareMathAlphabet\mathbfcal{OMS}{cmsy}{b}{n}

\newcommand{\mc}{\mathcal}
\newcommand{\bit}{\begin{itemize}}
\newcommand{\eit}{\end{itemize}}
\newcommand{\ben}{\begin{enumerate}} 
\newcommand{\een}{\end{enumerate}}
\newcommand{\bbm}{\begin{bmatrix}}
\newcommand{\ebm}{\end{bmatrix}}

\renewcommand{\l}{\left}
\renewcommand{\r}{\right}

\def\wh{\widehat}
\def\wt{\widetilde}

\newcommand{\E}[0]{\mathsf{E}}

\theoremstyle{definition}

\theoremstyle{definition}

\theoremstyle{definition}

\theoremstyle{definition}

\theoremstyle{definition}

\theoremstyle{remark}

\theoremstyle{definition}

\theoremstyle{definition}

\numberwithin{equation}{section}
\numberwithin{figure}{section}
\numberwithin{table}{section}

\title{Discussion of
\\
`Detecting possibly frequent change-points: Wild Binary Segmentation 2 and steepest-drop model selection'}

\author{Haeran Cho$^*$ and Claudia Kirch}

\begin{document}

\maketitle

We congratulate the author for this interesting paper \citep{fryzlewicz2018}
which introduces a novel method for the data segmentation problem 
that works well in a classical change point setting as well as in a frequent jump situation. 
Most notably, the paper introduces a new model selection step 
based on finding the `steepest drop to low levels' (SDLL). 
Since the new model selection requires a complete (or at least relatively deep) solution path 
ordering the change point candidates according to some measure of importance, 
a new recursive variant of the Wild Binary Segmentation \cite[WBS]{fryzlewicz2014}
named WBS2, has been proposed for candidate generation.

\section{Theoretical properties}
\label{section_2}

One of the main strengths of the proposed methodology, possibly due to the SDLL, 
is that it can work well both in a {\it change point} regime as well as in a {\it frequent jump} regime: In a 
change point regime the minimum distance to the next change,
%$\delta_T := \min_{1 \le i \le N + 1} (\eta_i - \eta_{i - 1})$,
$\delta_i:=\min(\eta_i-\eta_{i-1},\eta_{i+1}-\eta_i)$,
is reasonably large while the magnitude of the change $f^\prime_i$ is bounded from above 
and can be small (even tend to zero as $T \to \infty$).
In a frequent jump regime $\delta_i$ is small (related to outlier detection) 
and necessarily corresponding jumps $f_i^{\prime}$ need to be large to be detectable. 
In both situations, an adaptation of Lemma~1 of \cite{wang2018d} shows that
no consistent estimator of the locations of change point exists 
when $\sigma^{-2} \min_{1 \le i \le N} (\delta_i\,(f_i^{\prime})^2) < \log(T)$.

While WBS2.SDLL is shown to perform well in both regimes numerically,
the paper does not provide a theoretical underpinning of this good behaviour, 
in the sense that only a linear-time change point setting with 
$\delta_T:=\min_i\delta_i$ being of the same order as the sample size $T$ is considered:
Such an assumption is not necessary for consistent change point detection and, moreover,
it excludes models such as {\tt extreme.teeth} (ET) and {\tt extreme.extreme.teeth} (EET),
which are reasonably considered as belonging to the frequent jump regime
with $\delta_T \le 5$.
In the future, it will be very exciting to see which theoretical framework 
will help us to better understand the performance of statistical procedures
that aim at handling both regimes simultaneously.

In addition, the best currently available results for the localisation rate attained by WBS
as well as the requirement on the magnitude of changes for their detection,
are sub-optimal when $\delta_T/T \to 0$
(see Proposition~3.4 of \cite{ck2019}).
\cite{baranowski2016} and \cite{wang2018d} suggest modifications of WBS that 
alleviate the sub-optimality at the cost of introducing additional tuning parameters
such as a threshold or an upper bound on the length of random intervals.
However, even in these papers, the assumptions are formulated in terms of 
$ \min_i\delta_i\,\min_i (f_i^{\prime})^2$,
which does not reflect that the strength of multiscale procedures lies
in their ability to handle data sets containing both small changes with long distances to neighbouring change points,
as well as large changes with shorter distances (see e.g., the {\tt mix} model).
\cite{ck2019} consider multiscale change point situations
by working with $\min_i (\delta_i\,(f_i^{\prime})^2) $ 
in the theoretical investigation of a more systematic moving sum (MOSUM)-type procedure
for candidate generation.

\section{SDLL with alternative candidate generation methods}

As already pointed out by the author, both components of the proposed algorithm, 
i.e.,\ candidate generation and model section, can be used in combination with other methods. 
For example, in \cite{ck2019}, a version of WBS2 has been adopted 
as a candidate generation method for the localised pruning method proposed for model selection.
We will now show that deterministic candidate generation methods,
such as the multiscale MOSUM procedure \citep{chan2017, ck2019},
can be used with SDLL. 
Our first tentative attempt at generating a complete solution path of candidates
with a reasonable measure of importance attached, is described in Section~\ref{section_3} below.
Based on some initial simulation results reported in Table~\ref{table:one},
we conclude that deterministic candidate generation methods can be a good alternative, 
and that this approach merits further research.
Such a deterministic method will always yield the same result when applied to the same data set,
whereas WBS-based methods can produce different outcomes in different runs
(as observed in \cite{ck2019} on array comparative genomic hybridization data sets).
In particular, WBS-based results are reproducible only if the seed of the random number generation is also reported. 
In Section~4.1 of the present paper, the use of  a `median' of several runs is proposed 
to reduce this problem, which clearly comes at the cost of additional computation time.

\section{MOSUM-candidate generation and some simulations}
\label{section_3}

Many of the methods included in the comparative simulation studies of the present paper
have been designed for the change point regime with their default parameters chosen accordingly,
e.g., to save computation time. 
For example, the algorithm referred to as `MOSUM' in the present paper,
implemented in the R package {\tt mosum} \citep{mosum},
has a tuning parameter that relates to the smallest $\delta_T$ permitted,
and its default value is set at $10$, which we consider as a reasonable lower bound for a change point problem.
Also, the default choice of the parameter $\alpha \in [0, 1]$,
which stems from change point testing and sets a threshold for candidate generation in the algorithm,
is somewhat conservative ($\alpha = 0.1$) and not very meaningful in the frequent jump regime. 
Moving away from the change point regime,
we set the minimum bandwidth as small as possible in generating the bandwidth set $\mc G$,
\footnote{We generate $\mc G$ as detailed in Section~3.5 of \cite{meier2018} with $G_1 = 1$ 
but only use bandwidths $\ge 2$ due to the necessity of local variance estimation.}
and also set a more liberal threshold with $\alpha = 0.9$.
With these choices, MOSUM shows much better performance than that reported in the present paper,
see Table~\ref{table:one} below. 

Additionally, we explore the possibility of deterministic candidate generation 
based on moving sum statistics for a given set of bandwidth pairs $(G_l, G_r) \in \mc G \times \mc G$:
\begin{align*}
\wt M_k(G_l, G_r; X) = \sqrt{\frac{G_l G_r}{G_l + G_r}} \l(\frac{1}{G_l}\sum_{t = k - G_l + 1}^k X_t
- \frac{1}{G_r} \sum_{t = k + 1}^{k + G_r} X_t \r).
\end{align*}
At each scale $(G_l, G_r)$, we identify all $\wt k$ which maximises 
$\vert \wt M_k(G_l, G_r; X) \vert$ locally within $(\wt k- G_l, \wt k + G_r)$,
denote the collection of such $\wt k$ by $\mc K(G_l, G_r)$,
and set $M_k(G_l, G_r; X) = \wt M_k(G_l, G_r; X) \cdot 
\mathbb{I}\{k \in (\wt k - G_l, \wt k + G_r), \, \wt k \in \mc K(G_l, G_r)\}$.
We aggregate the MOSUM statistics generated at multiple scales as
$V(k) = \sum_{(G_l, G_r)} M_k(G_l, G_r; X)$, and then generate a solution path as in Algorithm~\ref{alg:one},
which is complete if the scale $(1, 1)$ is included.

\begin{algorithm}[ht]
\caption{MOSUM-based solution path generation}
\label{alg:one}
\DontPrintSemicolon
\SetAlgoLined
\SetKwFunction{exhaustive}{{\tt PrunAlg}}

% \KwIn{Data $\{X_t\}_{t = 1}^T$}
\BlankLine
Set the initial solution path $\mc P = \emptyset$

\Repeat{$V(k) = 0 \, \forall \, k$}{
	Find $k^\circ \leftarrow \arg\max_{1 \le k < T} \vert V(k) \vert$ 
	and $(G_l^\circ, G_r^\circ) \leftarrow \arg\max_{(G_l, G_r)} \vert M_{k^\circ}(G_l, G_r) \vert$
	\BlankLine
	
	Add % $(k^\circ, G_l^\circ, G_r^\circ, \vert M_{k^\circ}(G_l^\circ, G_r^\circ) \vert)$ 
	$(k^\circ, \vert M_{k^\circ}(G_l^\circ, G_r^\circ) \vert)$ to $\mc P$
	\BlankLine
	
	For each $(G_l, G_r)$, set $M_k(G_l, G_r) \leftarrow 0$
	% for any $k \in [k^\circ - \lfloor \eta G_r \rfloor, k^\circ + \lfloor \eta G_l \rfloor]$
	for any $k \in (\wt k - G_l, \wt k + G_r)$
	with $\wt k \in \mc K(G_l, G_r) \cap (k^\circ - G_r, k^\circ + G_l)$
	\BlankLine
	
	Update $V(k) \leftarrow \sum_{(G_l, G_r)} M_k(G_l, G_r; X)$
}
\BlankLine
\KwOut{$\mc P$}
\end{algorithm}

Referring to the methodology combining Algorithm~\ref{alg:one} with SDLL
as MOSUM.SDLL, Table~\ref{table:one} shows the results from applying WBS2.SDLL, MOSUM.SDLL
(both with $\lambda = 0.9$) and MOSUM
(with the aforementioned choice of parameters) to ET and EET
summarised over $1000$ realisations.
All methods perform better for EET than for ET
since the signal-to-noise ratio $\sigma^{-2} \min_i \delta_i\,(f_i^\prime)^2$  
is greater for ET (see also Section~\ref{section_2} above). 

As already mentioned, MOSUM adapted for the frequent jump regime works considerably better 
than the default version calibrated for the change point regime. 
While being more conservative than the SDLL-based methods for ET,
MOSUM still outperforms the others in terms of the absolute and the squared error measures
and overall, it returns reasonably good estimators at a fraction of the time.
MOSUM.SDLL shows that the deterministic candidate generation
provides a promising alternative to WBS2: It performs slightly worse than WBS2.SDLL
in identifying the correct number of change points ($N = 199$)
but the mean squared error of $\wh f$ indicates that MOSUM.SDLL may return estimators of better localisation accuracy.

\begin{table}[htb]
	\caption{Simulations results as in Table~2 of \cite{fryzlewicz2018}.}\label{table:one}
\resizebox{\columnwidth}{!}
{\small
\begin{tabular}{c|ccc|ccc}
\hline\hline
&	\multicolumn{3}{c}{{\tt extreme.teeth}} &			\multicolumn{3}{c}{{\tt extreme.extreme.teeth}} 			\\	
&	WBS2.SDLL &	MOSUM &	MOSUM.SDLL &	WBS2.SDLL &	MOSUM &	MOSUM.SDLL 	\\	\hline
$\wh{\E}(\wh N - N)$ &	 0.312 &	$- 3.002$ &	$- 1.689$ &	 0.264 &	 0.261 &	0.153	\\	
$\wh{\E}|\wh N - N|$ &	 3.628 &	 3.440 &	 5.345 &	 0.766 &	 1.107 &	1.047	\\	
$\wh{\E}(\wh N - N)^2$ &	 25.776 &	 19.798 &	 52.407 &	 1.896 &	 2.493 &	2.861	\\	
$\wh{\E}(\wh f - f)^2$ &	 0.049 &	 0.049 &	 0.041 &	 0.017 &	 0.017 &	0.016	\\	
time &	 0.180 &	 0.067 &	 0.770 &	 0.128 &	 0.055 &	0.481	\\	
\hline
\end{tabular}}
\end{table}

\small
\bibliographystyle{asa}
\bibliography{fbib}

\end{document}